\documentclass{ws-ijmpcs}

\begin{document}

\markboth{J. M. Alarc\'on}
{The role of the $\Delta(1232)$-resonance in
covariant baryon chiral perturbation theory}

%
\catchline{}{}{}{}{}
%

\title{The role of the $\Delta(1232)$-resonance in
covariant baryon chiral perturbation theory}

\author{J. M. Alarc\'on}

\address{PRISMA Cluster of Excellence, Johannes Gutenberg-Universit\"at, Mainz D-55099, Germany\\
Institut f\"ur Kernphysik, Johannes Gutenberg-Universit\"at,
  Mainz D-55099, Germany\\
alarcon@kph.uni-mainz.de}

\maketitle

\begin{history}
\received{Day Month Year}
\revised{Day Month Year}
\end{history}

\begin{abstract}
We stress, on theoretical and phenomenological grounds, the importance of the $\Delta(1232)$-resonance in a chiral effective field theory approach applied to the study of $\pi N$ scattering. We show how its inclusion as a dynamical degree of freedom allow us to obtain reliably valuable information from $\pi N$ scattering data.

\keywords{$\Delta(1232)$ resonance; $\pi N$ scattering; Baryon chiral perturbation theory; $\pi N$ sigma term.}
\end{abstract}

\ccode{PACS numbers: 13.75.Gx, 11.30.Rd, 12.39.Fe, 13.85.Dz}

\section{Introduction}	

The role of the $\Delta(1232)$-resonance in baryon chiral perturbation theory (BChPT) has been an old and interesting topic since the beginning of the 90's\cite{Jenkins:1991es}, and
its importance is related to the proper formulation of the effective field theory. 
This is because the $\Delta(1232)$ couples strongly to the $\pi N$ system already at low energies, what translates into a strong influence of this resonance in the $\pi N$ scattering process even at very low energies.
This motivates, from the phenomenological point of view, the explicit inclusion of this resonance, since one can expect that it will play an important role as an active degree of freedom in an effective field theory of $\pi N$ scattering.
Through this inclusion, one keeps a proper separation of scales and can extend the range of validity of the perturbative treatment.\cite{Alarcon:2012kn}
On the other hand, one can also justify this inclusion invoking large $N_c$ limit arguments since, in the emerging spin-flavor symmetry, nucleons and deltas are part of the same mutliplet.\cite{CalleCordon:2012xz}


\section{Extending BChPT}	

The possible ways to incorporate the $\Delta(1232)$ in chiral effective field theory (EFT) have been the subject of intensive research. Some crucial questions that one must address are: 1) How to filter the unphysical degrees of freedom introduced by the Rarita-Schwinger (RS) field that represents the $\Delta$, and 2) how to deal with the new scale $m_\Delta$, i. e., redefine a counting that takes into account the mass splitting $\delta=m_\Delta-m_N$. For the results that we will show here, we employed the consistent formulation of chiral Lagrangians\cite{Pascalutsa:1999zz} in combination with the $\delta$-counting.\cite{Pascalutsa:1999zz,Hanhart:2002bu} 
Another issue related to chiral effective field theory with baryons is how to keep explicitly Lorentz invariance in the formulation without breaking the standard chiral counting.
Here we employ the extended-on-mass-shell scheme\cite{EOMSrefs} (EOMS), which will be an important ingredient in the correct extraction of the $\pi N$ scattering observables.\cite{Alarcon:2012kn} 
This allow us to overcome some problems that the Heavy Baryon (HBChPT) and Infrared Regularization (IR) formulations have in extracting some important information related to $\pi N$ scattering phenomenology.
In the following we will show how the inclusion of the $\Delta$ in the EOMS renormalized BChPT allow us to exploit the theory to obtain valuable information directly from experimental data. 
This EOMS renormalized approach is what we call covariant BChPT.

\section{$\pi N$ scattering phenomenology}\label{Sec:piNscatteringphenomenolgy}

In this section we summarize some important results obtained from the inclusion of the $\Delta$ in  covariant BChPT.
The crucial improvement achieved by this approach is directly related to the convergence of the chiral series, which translates into an improved description of the physical and the subthreshold regions.
Also, we obtained a better convergence in the chiral expansion of the extracted quantities related to $\pi N$ scattering.\cite{Alarcon:2012kn} 
Such improved convergence is not observed, however, in the IR formulation\cite{Alarcon:2012kn,Alarcon:2011kh}.
Notice that the lack of convergence in the subthreshold region in previous analyses questioned the applicability of BChPT to $\pi N$ scattering.\cite{Becher:2001hv} 

In the following we concentrate in two practical applications of our calculation: The extraction of the pion-nucleon sigma term ($\sigma_{\pi N}$) from $\pi N$ scattering data and its impact on the calculation of the strangeness content of the nucleon ($\sigma_s$).

\subsection{The pion-nucleon sigma term}\label{Sec:SigmaTerm}

The pion-nucleon sigma term is a quantity that contains information about the internal scalar structure of the nucleon which is used in a broad variety of nuclear-physics calculations.\cite{NuclearMatter,Antropic,Lacour:2010ci} 
However, it has become more known due to its impact on estimations of direct detection of dark matter\cite{DMDetection}, since it is related to the scalar coupling of the nucleon:

\begin{align}
\sigma_{\pi N}=\frac{\hat{m}}{2m_N} \langle N |  \bar{u}u +\bar{d}d  |N \rangle, \hspace{2cm}  \hat{m}=\frac{m_u+m_d}{2}.
\end{align}

Since this quantity has a direct impact on $\pi N$ scattering, it is possible to extract it in a model-independent way from a BChPT calculation of this process. 
This has been already attempted in previous works, but the lack of convergence did not allow to give a reliable estimation of $\sigma_{\pi N}$.
However, this problem can be overcome in the covariant approach of BChPT including the $\Delta$-resonance.\cite{Alarcon:2012kn,Alarcon:2011zs}
In this way one is able to extract a reliable value for $\sigma_{\pi N}$ employing only experimental information. 
We report, from modern $\pi N$ scattering data, a value of\cite{Alarcon:2012kn,Alarcon:2011zs} 

\begin{align}
\sigma_{\pi N}=59(7)\text{~MeV.}
\end{align}

This result, on the other hand, is compatible with updated $\pi N$-scattering phenomenology and modern pionic atoms data\cite{Alarcon:2011zs}, and also with a recent calculation based on $N_f=2+1$ LQCD world data.\cite{Alvarez-Ruso:2013fza}

\subsection{The strangeness content of the nucleon}\label{Sec:Sigmas}

The value of $\sigma_{\pi N}$ deduced from scattering data has a direct impact on the strangeness content of the nucleon, defined as 

\begin{align}
\sigma_s\equiv \frac{m_s}{2 m_N}\langle N| \bar{s}s |N\rangle,
\end{align}

since the deviation of $\sigma_{\pi N}$ from the non-singlet contribution to the nucleon mass 

\begin{align}
\sigma_0\equiv \frac{\hat{m}}{2 m_N}\langle N|  \bar{u}u+\bar{d}d - 2 \bar{s}s |N \rangle
\end{align}

is proportional to $\sigma_s$.\cite{Gasser:1980sb,Alarcon:2012nr} 
The calculation of $\sigma_s$ through $\sigma_0$ has the important advantage that the latter can be obtained from the octet mass splittings.\cite{Gasser:1980sb}
In fact, with this method Gasser extracted from the hadron spectrum a value of $\sigma_0=35(5)$~MeV using a pre-ChPT model for the interaction of the baryon with the meson cloud.\cite{Gasser:1980sb}
On the other hand, a later calculation in $SU(3)$ HBChPT obtained a value of $\sigma_0=36(7)$~MeV.\cite{Borasoy:1996bx}
However, it is important to take into account that HBChPT has a poor convergence in the $SU(3)$ sector and also that the decuplet can play an important role in the final value of $\sigma_0$.\cite{Geng:2008mf}
Notice the importance of an accurate determination of this quantity in the final evaluation of  $\sigma_s$ through $\sigma_0$.\cite{Alarcon:2012nr}
For this reason, we updated the value of $\sigma_0$ with a covariant BChPT calculation including the contribution of the decuplet. 
We obtained a result of \cite{Alarcon:2012nr} 

\begin{align}
\sigma_{0}=58(8)~\text{MeV},
\end{align}

which implies,

\begin{align}
\sigma_{s}=16(80)~\text{MeV},
\end{align}

 for a $\sigma_{\pi N}=59(7)$~MeV.
This result, obtained uniquely from experimental information, is in good agreement with the LQCD estimations for the strangeness in the nucleon, that also point to a small content.\cite{Ohki:2008ff,Junnarkar:2013ac}
Such a good agreement is not obtained, however, from the classical value of $\sigma_{\pi N}=45(8)$~MeV,\cite{Gasser:1990ce} from where one obtains $\sigma_s=-150(80)$~MeV.\cite{Alarcon:2012nr}

\section{Summary and Conclusions}	

Based on our analyses of $\pi N$ scattering with and without the $\Delta(1232)$ included explicitly as a degree of freedom in BChPT within the IR and EOMS schemes\cite{Alarcon:2012kn,Alarcon:2011kh}, we analyzed which framework exploits better the possibilities of the theory to study the $\pi N$-scattering process. 
We found that, in order to achieve the best convergence properties, it is crucial to combine a consistent inclusion of the $\Delta(1232)$ resonance together with a renormalization scheme that preserves the analytical properties of a covariant calculation (EOMS).
Once this is done, the calculated chiral amplitude shows an improved convergence that allow us to extract reliably the $\pi N$ scattering phenomenology at low energies and below the threshold.\cite{Alarcon:2012kn}
This better convergence allowed us to solve previous disagreements with dispersive methods that questioned the applicability of BChPT to the study of $\pi N$ scattering.\cite{Becher:2001hv}
Also, as commented in Sec.~\ref{Sec:piNscatteringphenomenolgy}, this improvement allowed us to extract reliably and accurately the value of $\sigma_{\pi N}$ and $\sigma_s$ employing only experimental information. These were found to be compatible with updated phenomenology\cite{Alarcon:2011zs} and recent LQCD determinations.\cite{Alvarez-Ruso:2013fza,Ohki:2008ff,Junnarkar:2013ac}
We can conclude then, that the inclusion of the $\Delta(1232)$ resonance in covariant BChPT is very important to exploit the potential of chiral effective field theory with baryons.


\end{document}